\documentclass[%
 reprint,
 amsmath,amssymb,
 aps,
]{revtex4-2}
\usepackage{amssymb}
\usepackage{graphicx}
\usepackage{dcolumn}
\usepackage{bm}
\usepackage{float}
\usepackage{bm}
\usepackage{physics} 
\usepackage{xcolor} 
\newcommand{\CK}[1]{\textcolor{black}{#1}}

\begin{document}

\preprint{APS/123-QED}

\title{Quantum sensing through bosonic–fermionic Bell-state transitions in two-photon interference}

\author{Chahat Kaushik*}
\affiliation{%
 Photonic Sciences Lab., Physical Research Laboratory, Ahmedabad 380009, Gujarat, India.\\
}%
 \altaffiliation[Also at ]{Indian Institute of Technology-Gandhinagar,  382055, Gujarat, India.}
  \email{kaushik.28chahat@gmail.com}
\author{Vimlesh Kumar}%
\affiliation{%
 Photonic Sciences Lab., Physical Research Laboratory, Ahmedabad 380009, Gujarat, India.\\
}%

\author{G. K. Samanta}%

\affiliation{%
 Photonic Sciences Lab., Physical Research Laboratory, Ahmedabad 380009, Gujarat, India.\\
}%

\date{\today}

\begin{abstract}
Hong–Ou–Mandel (HOM) interference has become a central resource for quantum sensing and metrology owing to its sensitivity to temporal delay and photon indistinguishability. However, existing HOM-based sensing schemes generally rely on inserting a sample into one arm of the interferometer, making the measurement vulnerable to optical loss, alignment instability, and bandwidth-dependent distortion of the interference profile. Here, we demonstrate a symmetry-controlled quantum sensing scheme based on continuous transitions between symmetric (bosonic-like) and antisymmetric (fermionic-like) Bell states in two-photon interference. By imprinting a geometric phase onto the classical pump beam and transferring it to polarization-entangled photons generated via spontaneous parametric down-conversion, we coherently tune the exchange symmetry of the entangled state without altering the temporal or spectral indistinguishability of the photons. The HOM response evolves continuously from bunching to antibunching with a sine square phase dependence, producing a coincidence modulation of $\sim10\times10^4$ counts s$^{-1}$. In contrast to conventional HOM sensing, the phase-modulation linewidth remains fixed at $\pi/2$, independent of photon bandwidth. Using a birefringent crystal placed directly in the pump beam, we measure thermo-dispersive birefringence with a resolution of the order of $10^{-6}$ over a broad temperature range. Our results establish exchange symmetry as a controllable resource for robust quantum sensing and symmetry-engineered photonic quantum information processing.


\end{abstract}

\maketitle


\section{\label{sec:level1}Introduction}

Quantum interference of indistinguishable photons lies at the heart of quantum information science, quantum metrology, and photonic quantum technologies \cite{Zeilinger:2000, Shih:1988, Knill:2001, Branning:1999, Nielsen:2010, Giovannetti:2004}. Among different manifestations of quantum interference, Hong–Ou–Mandel (HOM) interference \cite{HoM1987} has emerged as one of the most fundamental and experimentally accessible phenomena for probing indistinguishability and exchange symmetry of photons \cite{Branning:1999, Pittman:1996}. When two indistinguishable photons simultaneously interfere at a balanced 50:50 beam splitter, due to bosonic exchange symmetry, both photons bunch together, producing a characteristic suppression of coincidence counts known as the HOM dip. Owing to its extreme sensitivity to temporal delay, spectral overlap, polarization, and phase coherence, HOM interferometry has evolved into a versatile platform for quantum sensing \cite{Singh:24, Johnson:2023}, quantum imaging \cite{Ndagano:2022, Johnson:2023}, optical delay measurements \cite{Lyons:2018, Singh:2023b}, clock synchronization \cite{Liu:2021}, and characterization of photonic quantum states \cite{White:2025}.

Conventional HOM-based sensing schemes generally rely on introducing a sample into one arm of the interferometer \cite{Ndagano:2022, Singh:24, Johnson:2023}, where the physical parameter of interest modifies the relative optical delay or spectral phase between the two photons and is estimated through coincidence detection. Such approaches have demonstrated remarkable sensitivity beyond classical interferometric techniques because the coincidence probability depends directly on two-photon interference rather than on single-photon intensity modulation. However, the sensitivity and resolution of these schemes are strongly influenced by the bandwidth and coherence properties of the photon pairs \cite{Singh:2023b, Okano:2015, Chen:2019}. In addition, placing the sample inside the interferometer introduces several practical limitations, including optical loss, reduced coincidence rates, alignment instability, dispersion-induced asymmetry of the HOM profile, and degradation of interference visibility. These effects become increasingly significant in precision measurements involving weak signals or long optical interaction lengths.

The quantum interference can also be understood from the perspective of exchange symmetry of multiphoton states \cite{Branning:1999, Liu:2022, Gao:2022, Mei:2026}. In particular, polarization-entangled Bell states provide a natural framework for exploring symmetric and antisymmetric exchange properties of two-photon systems. While symmetric Bell states exhibit bosonic-like bunching in HOM interference, antisymmetric Bell states behave analogously to fermionic particles and produce antibunching. Intermediate phase-shifted Bell states can further emulate anyonic-like interference behavior through controlled superposition of symmetric and antisymmetric exchange amplitudes or unsymmetrized states \cite{Branning:1999}. Therefore, the control over exchange symmetry, achieved through the relative phase between the two terms of the Bell states, provides an additional degree of freedom for manipulating quantum interference beyond conventional temporal-delay engineering. The relative phase between the two terms of the Bell states can arise from the different sources, including the relative phase of the pump \cite{Jabir:2017direct, Qi:2020, kumar:2025}, birefringence of the nonlinear crystal producing the pair photons \cite{Kwiat:1995, Rangarajan:2009, Davis:2025} are detrimental and requires employment of phase compensation schemes \cite{Altepeter:2005, Rangarajan:2009, Davis:2025}, for real world applications of the Bell states. However, in recent times, it has been observed that the unwanted relative phase of the Bell states has been used as the controlled parameter by transferring the geometric phase \cite{Pancharatnam1956, Berry1984} of the pump beam \cite{Qi:2020, kumar:2025} for quantum state manipulation \cite{kumar:2025}, phase memory \cite{Qi:2020}, and phase compensation in quantum key distribution \cite{nai2026geometric}. Unlike the dynamic phase, the geometric phase depends only on the evolution path of the polarization state, providing a stable and loss-insensitive control of entangled-photon interference. Although geometric phases have been extensively studied in classical optics, structured light, and quantum state evolution, their role in continuously controlling exchange symmetry and exploiting it for quantum sensing remains largely unexplored \cite{Liu2022}.

Here, we demonstrate a symmetry-controlled quantum sensing scheme based on continuous transitions between symmetric (bosonic-like) and antisymmetric (fermionic-like) Bell states in HOM interference. By introducing a controllable geometric phase between the orthogonal polarization components of a classical pump beam and transferring it to polarization-entangled photons generated through spontaneous parametric down-conversion, we coherently tune the exchange symmetry of the entangled state without modifying the optical delay, spectral bandwidth, or temporal indistinguishability of the photons. The HOM interference evolves continuously from bunching to antibunching with coincidence modulation following a $\sin^2(\phi)$ dependence. Unlike conventional HOM sensing, the phase-modulation linewidth remains independent of photon bandwidth, enabling robust phase-sensitive measurements. Using this approach, we demonstrate high-resolution thermo-dispersive birefringence measurements by placing the sample directly in the classical pump beam instead of inside the interferometer. Our results establish exchange symmetry as a controllable resource for quantum sensing and provide a new route toward symmetry-engineered photonic quantum information processing.

\section{Theory: Exchange-Symmetry of Bell states in HOM interferometry}

\noindent To illustrate the effect of relative phase on entangled photon states, we consider a general polarization-entangled state of two photons of the form
\begin{equation}
    |\psi_\phi\rangle= \frac{1}{\sqrt{2}}(|HH\rangle\pm e^{i2\phi} |VV\rangle)
    \label{eq1}
\end{equation}
where $2\phi$ is the relative phase between the two terms of the Bell state introduced by the geometric phase or any phase object between the horizontal and vertical polarization components. Now, if we put a half-wave plate (HWP1) with fast axis oriented at an arbitrary angle $\delta$ with respect to horizontal, the state of the photon passing through it changes as

\begin{equation}
\begin{split}
 |H\rangle=\cos(2\delta)|H\rangle+\sin(2\delta)|V\rangle)\\
 |V\rangle=\sin(2\delta)|H\rangle-\cos(2\delta)|V\rangle)
\end{split}
\label{eq2}
\end{equation}
and the polarization-entangled state given by Eq. \ref{eq1} is transformed into 

\begin{equation}
\begin{aligned}
|\psi_{\phi,\delta}\rangle
&= \frac{1}{\sqrt{2}} \Big[
\cos(2\delta)\,|HH\rangle
+ \sin(2\delta)\,|HV\rangle \\
&\quad \pm e^{i2\phi}\sin(2\delta)\,|VH\rangle
\mp e^{i2\phi}\cos(2\delta)\,|VV\rangle
\Big] \\
&= \cos(2\delta)\,\frac{|HH\rangle \mp e^{i2\phi}|VV\rangle}{\sqrt{2}} \\
&\quad + \sin(2\delta)\,\frac{|HV\rangle \pm e^{i2\phi}|VH\rangle}{\sqrt{2}}
\label{eq3}
\end{aligned}
\end{equation}

The Eq. \ref{eq3} can be simplified, considering one sign of each term, in a full Bell-basis representation given as
\begin{equation}
    \begin{split}
        |\psi_{\phi,\delta}\rangle=\cos(2\delta)(\cos(\phi)|\Phi^-\rangle-i\sin(\phi)|\Phi^+\rangle)+\\\sin(2\delta)(\cos(\phi)|\psi^+\rangle-i\sin(\phi)|\psi^-\rangle)
    \end{split}
    \label{eq4}
\end{equation}
From Eq. \ref{eq4}, it is evident that using different values of $\delta$ and $\phi$ we can transform the final output state into one of the four Bell states. As shown in table \ref{tab:states}, the two-photon output states of Eq. \ref{eq4} transform into one of the four Bell states for different combinations of $\delta$ and $\phi$.

\begin{table}[h!]
\centering
\caption{Output states for different values of $\phi$ and $\delta$.}
\label{tab:states}
\begin{tabular}{c@{\hspace{1.2em}}|c@{\hspace{1.5em}}|c@{\hspace{1.5em}}|c}
\hline
$\mathbf{\delta}$ & $\mathbf{\phi}$ & \textbf{Output state $|\psi_{\phi,\delta}\rangle$} & \textbf{State type} \\
\hline
\hline

$0^\circ$ & $0^\circ$ 
& $|\Phi^{-}\rangle=\frac{1}{\sqrt{2}}(|HH\rangle - |VV\rangle)$ 
& Symmetric \\

$0^\circ$ & $90^\circ$ 
& $-i|\Phi^{+}\rangle=\frac{-i}{\sqrt{2}}(|HH\rangle + |VV\rangle)$ 
& Symmetric \\

$45^\circ$ & $0^\circ$ 
& $|\Psi^{+}\rangle=\frac{1}{\sqrt{2}}(|HV\rangle + |VH\rangle)$ 
& Symmetric \\

$45^\circ$ & $90^\circ$ 
& $-i|\Psi^{-}\rangle=\frac{-i}{\sqrt{2}}(|HV\rangle - |VH\rangle)$ 
& Anti-symmetric \\

\hline
\end{tabular}
\end{table}

If the pair photon state described by Eq. \ref{eq4} is incident on the two input ports, $a$ and $b$ of an unbiased $50{:}50$ beam splitter (BS), forming a Hong–Ou–Mandel (HOM) interferometer, using the beam-splitter transformation equations we can write the state at the output ports, $c$ and $d$, as

\begin{equation}
\begin{split}
|\psi_{\text{out}}\rangle =\;&
\frac{i\cos(2\delta)\cos\phi}{2}
 \Big[\Big(|2_H\rangle_c|0\rangle_d + |0\rangle_c|2_H\rangle_d\Big) \\
 &- \Big(|2_V\rangle_c|0\rangle_d + |0\rangle_c|2_V\rangle_d\Big)\Big]\\
&\frac{\cos(2\delta)\sin\phi}{2}\Big[\Big(|2_H\rangle_c|0\rangle_d + |0\rangle_c|2_H\rangle_d\Big) \\
&+ \Big(|2_V\rangle_c|0\rangle_d + |0\rangle_c|2_V\rangle_d\Big)\Big]
\\
&+ i\frac{\sin(2\delta)\cos\phi}{\sqrt{2}}
\Big(
|1_H 1_V\rangle_c|0\rangle_d + |0\rangle_c|1_H 1_V\rangle_d
\Big) \\
&- \frac{i\,\sin(2\delta)\sin\phi}{\sqrt{2}}
\Big(
|1_H\rangle_c|1_V\rangle_d - |1_V\rangle_c|1_H\rangle_d
\Big).
\end{split}
\label{eq5}
\end{equation}

It is evident from the above expression that the beam splitter (BS) projects the symmetric Bell states, $|\Phi^-\rangle$, $|\Phi^+\rangle$, and $|\Psi^+\rangle$, into bunched NOON-like states comprising photons with identical or orthogonal polarizations, while the antisymmetric state $|\Psi^-\rangle$ remains unchanged and gives rise to antibunching. The relative contributions of these output components are governed by the parameter $\delta$, whereas their amplitudes are controlled by the geometric phase $\phi$. Thus, by tuning $\delta$ and $\phi$, one can selectively generate desired NOON states or preserve the Bell state as required. Using Eqs. \ref{eq4} and \ref{eq5}, the coincidence probability at the output ports $c$ and $d$ of the HOM interferometer can be expressed as,

\begin{equation}
\begin{split}
 P_c(\tau,\phi)=\frac{1}{2}(\cos^2(2\delta)[1- |S(\tau)|^2])+\\\frac{1}{2}(\sin^2(2\delta)[1-\cos(2\phi) |S(\tau)|^2]   
\end{split}
\label{eq6}
\end{equation}

Here, $S(\tau)=\int dt\, f(t)f^*(t-\tau)$ is the temporal overlap function for indistinguishable photons. At zero temporal delay ($\tau = 0$), $S(\tau)=1$. In this case, the coincidence contribution corresponding to the Bell states $\Phi^+$ and $\Phi^-$ vanishes, leading to photon bunching independent of the relative phase $\phi$. In contrast, the coincidence contribution associated with the Bell states $\Psi^+$ and $\Psi^-$ depends sinusoidally on $\phi$. Consequently, the coincidence probability can be written as 

\begin{equation}
    P_c(\phi,\delta)=\sin^2(2\delta)\sin^2(\phi)
    \label{eq7}
\end{equation}

As evident from this expression, for $\delta = 45^\circ$, the coincidence probability varies continuously from a minimum (photon bunching) at $\phi=0$ (corresponding to $|\Psi^+\rangle$) to a maximum (anti-bunching) at $\phi=90^\circ$ (corresponding to $|\Psi^-\rangle$). Thus, the relative phase $\phi$ enables continuous control between symmetric (bosonic-like) and antisymmetric (fermionic-like) two-photon states, effectively parameterizing a transition between exchange eigenstates without altering the spectral or temporal indistinguishability of the photons. The resulting change in HOM coincidence counts from a minimum to a maximum makes the HOM interferometer, in principle, a sensitive probe for physical parameters that influence the relative phase between the orthogonal polarization components of the classical pump beam generating the entangled state. Unlike conventional HOM-based group-index measurements \cite{Singh:24}, where the sample is inserted into one arm of the interferometer to introduce a relative optical delay between the photons, in the present scheme, the sample is placed directly in the classical pump beam. Consequently, the sample modifies the phase transferred to the entangled state and thereby controls the symmetry of the two-photon state without changing the optical delay between the photons. 

\section{Experimental Set up}

\begin{figure}[H]
    \centering
    \includegraphics[width=1\linewidth]{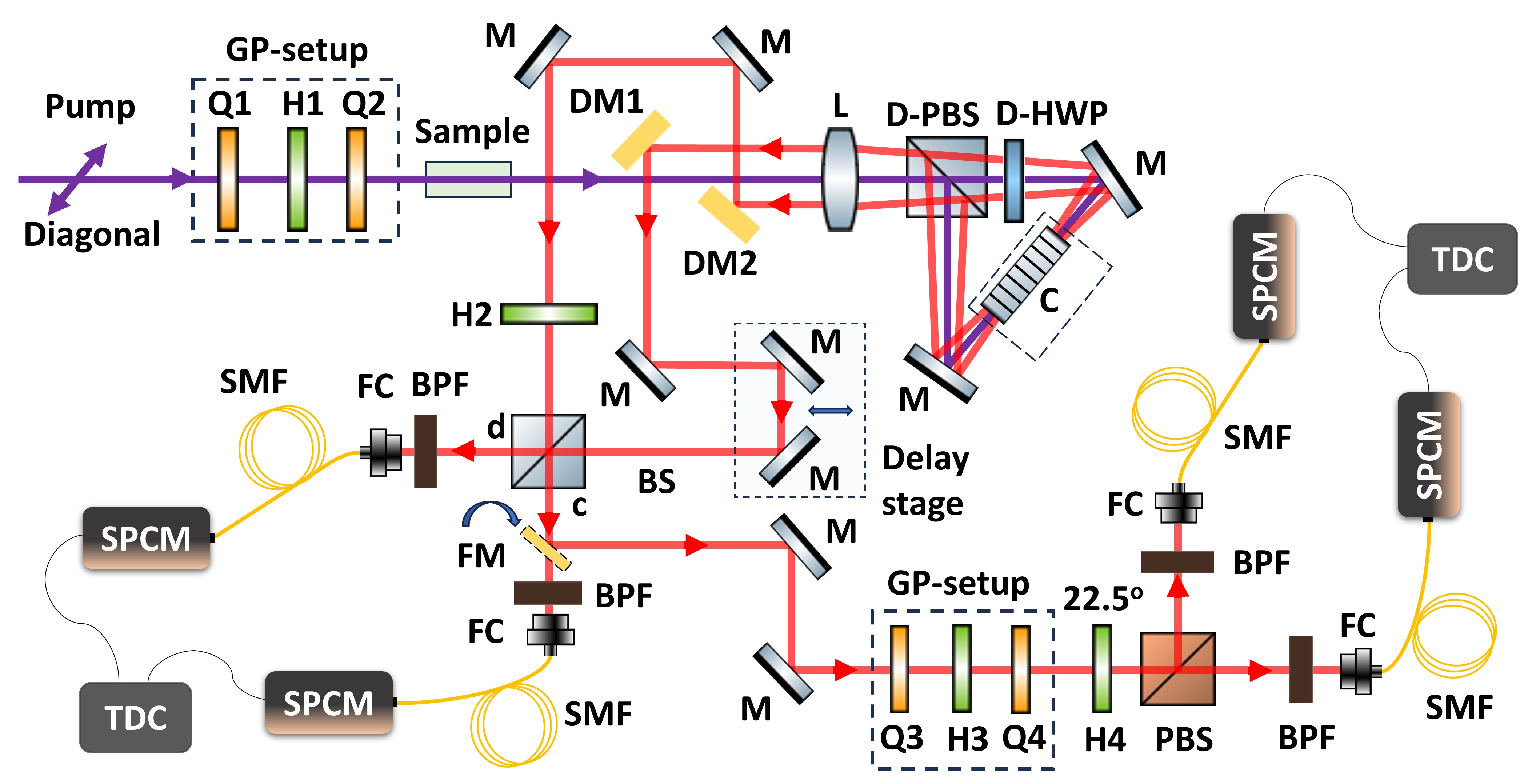}
    \caption{Schematic experimental setup for tunable exchange symmetry control of two photon state. Purple line denotes the pump laser beam at 405 nm, and red line denotes SPDC photons at 810 nm; Q1-4: quarter wave plates (QWP); H1-4: half wave-plates (HWP); GP-setup: geometric phase; M: mirror; D-PBS and D-HWP: dual wavelength (405 nm and 810 nm) polarizing beam-splitter cube and half-wave plate; C: PPKTP crystal; BS: beam splitter, DM1-2: D-shaped mirror; FC: Fiber coupler; SMF: single mode fibre; BPF: bandpass filter; SPCM: single photon counting module; TDC: time to digital converter; PBS: polarizing beam-splitter cube. Sample: PPKTP crystals of length, L= 5 mm and 10 mm.}
   \label{Figure 1}
\end{figure}

The schematic of the experimental setup for the HOM interference of an entangled state is shown in Fig. \ref{Figure 1}. A continuous-wave, single-frequency, linearly polarized laser diode at 405 nm, delivering up to 20 mW, is used as the pump source in the experiment. The geometric phase is imprinted to the pump beam by the GP setup \cite{Chahat:23, Chahat:25} comprised of a combination of a quarter wave-plate (Q1), half wave-plate (H1), and a second quarter wave-plate (Q2) with the fast axis at 45$^\circ$ (fixed), $\theta_p$ (variable), and 45$^\circ$ (fixed) with respect to vertical, respectively. The pump beam polarization to the GP setup is diagonal. To confirm the phase, the pump beam is characterized using the standard polarization-projection setup, which is not shown in Fig. \ref{Figure 1}. The sample is placed in the pump beam for the sensing experiment. To generate polarization-entangled pair photons, the pump beam is focused by a lens (L) of focal length $f = 150$ mm to the center of a 10-mm-long, $1 \times 1$ mm$^2$ periodically poled PPKTP crystal (C), placed symmetrically inside a polarization Sagnac interferometer. The interferometer consists of a dual-wavelength polarizing beam splitter (D-PBS), a dual-wavelength half-wave plate (D-HWP), and two high-reflecting mirrors for 405 nm and 810 nm, arranged such that the crystal center is at equal optical distance from the D-PBS for both clockwise (CW) and counterclockwise (CCW) pump paths. The PPKTP crystal, with a single grating of period $\Lambda$ = 3.425 $\mu$m, is maintained at $32.5^\circ$C with a stability of $\pm 0.1^\circ$C to achieve type-0 ($e \rightarrow e + e$), non-collinear, degenerate spontaneous parametric down-conversion (SPDC) at 810 nm. The pair photons generated in the CW and CCW directions are recombined at the D-PBS, resulting in a polarization-entangled state. The output, collimated by the lens, exhibits an annular intensity profile (observed using an EMCCD) with orthogonally polarized photon pairs. The pair photons are positioned across diametrically opposite points of the ring. The operating principle of such a Sagnac-based type-0 entangled photon source is described in Refs. \cite{Jabir:2017, Singh:2023}. 

%
Subsequently, the annular ring is split into two halves by the D-shaped mirrors (DM1, DM2) and guided using plane mirrors to the two input ports of the beam-splitter (BS) for HOM interference. While one of the arms of interferometer is delayed over other using the reflecting mirrors on the translation stage, the use of half-wave plate (H2) placed in one of the arms with fast axis at $45^\circ$ with respect to the vertical transforms the generated entangled photons in Bell state in the form of  $|\psi_{\phi}^-\rangle\rightarrow  \frac{1}{\sqrt{2}}(|H\rangle_s |H\rangle_i -e^{i2\phi} |V\rangle_s |V\rangle_i)$ to $|\psi_{\phi}^-\rangle\rightarrow \frac{1}{\sqrt{2}}(|H\rangle_s |V\rangle_i -e^{i2\phi} |V\rangle_s |H\rangle_i)$. The photons of the output ports of the BS are collected using the fiber collimator (FC) connected with a single-mode fiber (SMF) and detected using the single photon counting module (SPCM). The SPCM output is analyzed using a time-to-digital converter (TDC) to estimate quantum parameters in the experiment.  
Further, to analyze the output state of the HOM interferometer, photons from one of the output ports (here port c) are reflected by the flip mirror (FM) and a variable phase is introduced using the second geometric phase (GP) setup implemented using Q3, H3 and Q4 waveplates. The fast axes of the wave plates are oriented at $45^\circ$ (fixed), $\theta_a$ (variable), and $45^\circ$ (fixed) with respect to the vertical, respectively. The photons are then projected using a half-wave plate (H4) with its fast axis set at $22.5^\circ$, followed by a polarizing beam splitter (PBS). The resulting outputs are coupled into SMF through FC, detected using SPCMs, and analyzed via the TDC. In the experiment, the photons are extracted from the background using the band pass filters (BPF) of 3 nm centered at 810 nm). The coincidence window throughout the experiment is set to 2 ns unless otherwise specified.

\section{Results and Discussion}

\begin{figure}[h]
    \centering
    \includegraphics[width=\linewidth]{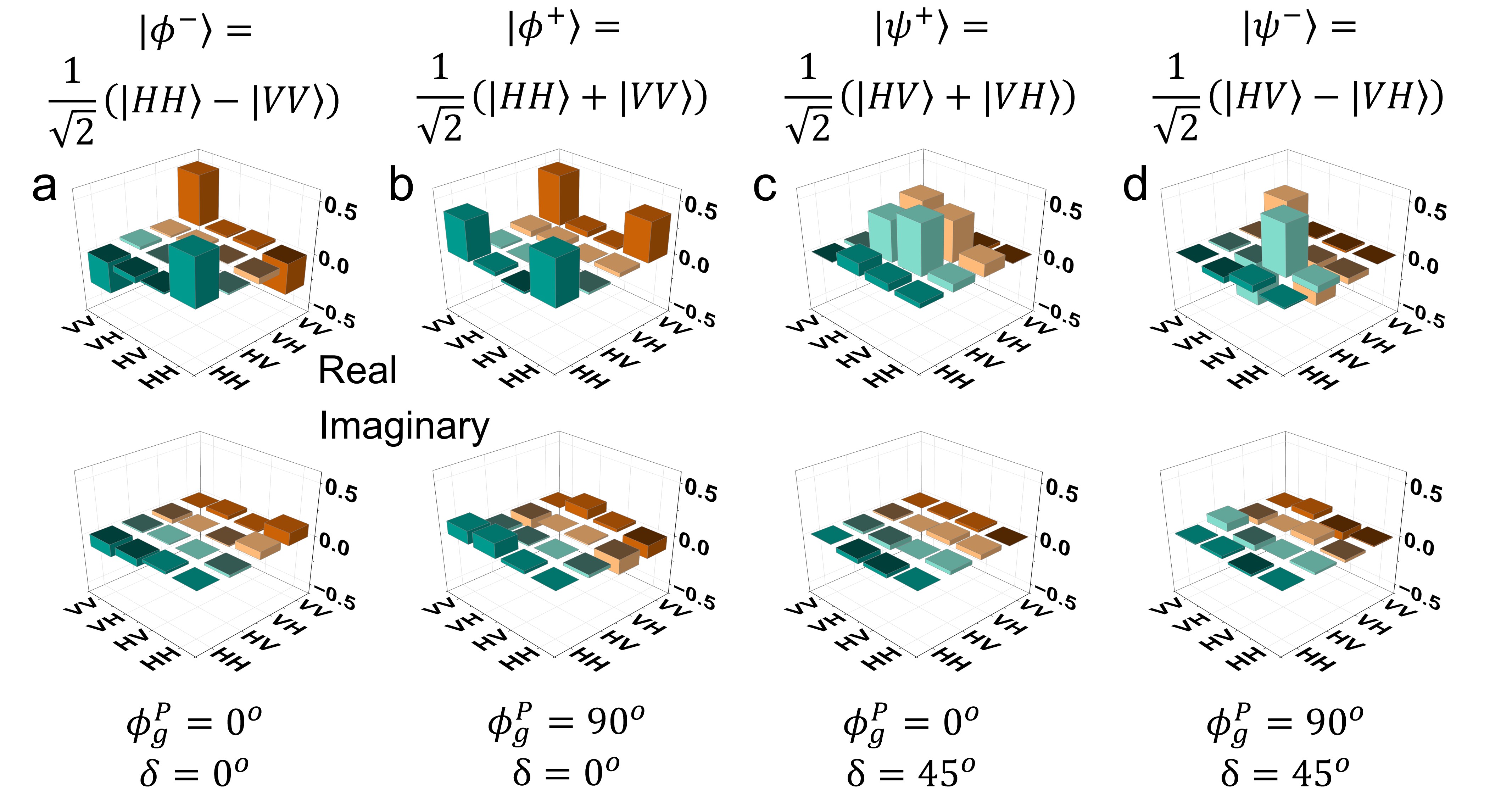}
    \caption{Real (top rows) and imaginary (bottom row) parts of the reconstructed density matrices for Bell states (a) $|\Phi^-\rangle$, (b) $|\Phi^+\rangle$, (c) $|\Psi^+\rangle$, (d) $|\Psi^-\rangle$.}
   \label{Figure 2}
\end{figure}

First, we verified the on-demand generation of Bell states described by Eq.~\ref{eq3} and summarized in Table~\ref{tab:states} using two controllable parameters: the geometric phase, $\phi_g^p$, and the angle $\delta$ of the HWP (H2), through quantum state tomography. Keeping the pump power fixed at 1 mW, we recorded coincidence counts for different polarization projection combinations and reconstructed the real and imaginary parts of the two-qubit density matrix for various values of the geometric phase, as shown in Fig. \ref{Figure 2}. From the reconstructed density matrices [real parts (top row) and imaginary parts (bottom row)] shown in Fig. \ref{Figure 2}(a-d), it is evident that for selected values of $(\phi_g^p,\delta)$ = $(0^\circ,0^\circ)$, $(90^\circ,0^\circ)$, $(0^\circ,45^\circ)$, and $(90^\circ,45^\circ)$, the generated states correspond to the four Bell states, $|\Phi^-\rangle$, $|\Phi^+\rangle$, $|\Psi^+\rangle$, and $|\Psi^-\rangle$, respectively. The corresponding state fidelities are 81\%, 87\%, 87\%, and 91\%. We further measured the polarization-correlation visibility of each Bell state in different measurement bases, obtaining average visibilities in the range of 94-98\%, with the CHSH Bell parameter, $S$, varying from 2.5-2.71.

All the measured parameters for the different Bell states are summarized in Table \ref{tab:bell_results}. Since the unitary transformations governed by $\phi_g^p$ and $\delta$ ideally generate different Bell states without altering the degree of entanglement, the visibility and CHSH-Bell parameter are expected to remain unchanged for all states. The small variations observed experimentally are primarily attributed to residual retardance and imperfections of the wave plates used in the setup. Furthermore, the scheme provides the flexibility to compensate unwanted relative phases in the Bell state arising from slight asymmetric positioning of the crystal inside the Sagnac interferometer or from the birefringence of the optical elements used in the experiment \cite{nai2026geometric} by appropriately introducing an offset geometric phase in the pump beam. 

\begin{table}[ht]
\centering
\caption{Measurement parameters of experimentally controlled Bell states.}
\renewcommand{\arraystretch}{1.15}
\setlength{\tabcolsep}{6pt}
\resizebox{\columnwidth}{!}{
\begin{tabular}{c|c|c|c|c}
\hline
\textbf{($\phi_g^p$, $\delta$)} & \textbf{Bell State} & \textbf{Avg. Vis. (\%)} & \textbf{S Value} & \textbf{Fidelity, F (\%)} \\
\hline
\hline
 0$^\circ$, 0$^\circ$ & $|\phi^-\rangle$ & $97.93$ & $2.519$ & $81$ \\
\hline
90$^\circ$, 0$^\circ$ & $|\phi^+\rangle$ & $97.3$ & $2.71$ & $87$ \\
\hline
0$^\circ$, 45$^\circ$ & $|\psi^+\rangle$ & $94.55 $ & $2.624$ & $87$ \\
\hline
90$^\circ$, 45$^\circ$ & $|\psi^-\rangle$ & $94.23 $ & $2.585$ & $91$ \\
\hline
\end{tabular}
}
\label{tab:bell_results}
\end{table}

\begin{figure}[h]
    \centering
    \includegraphics[width=1\linewidth]{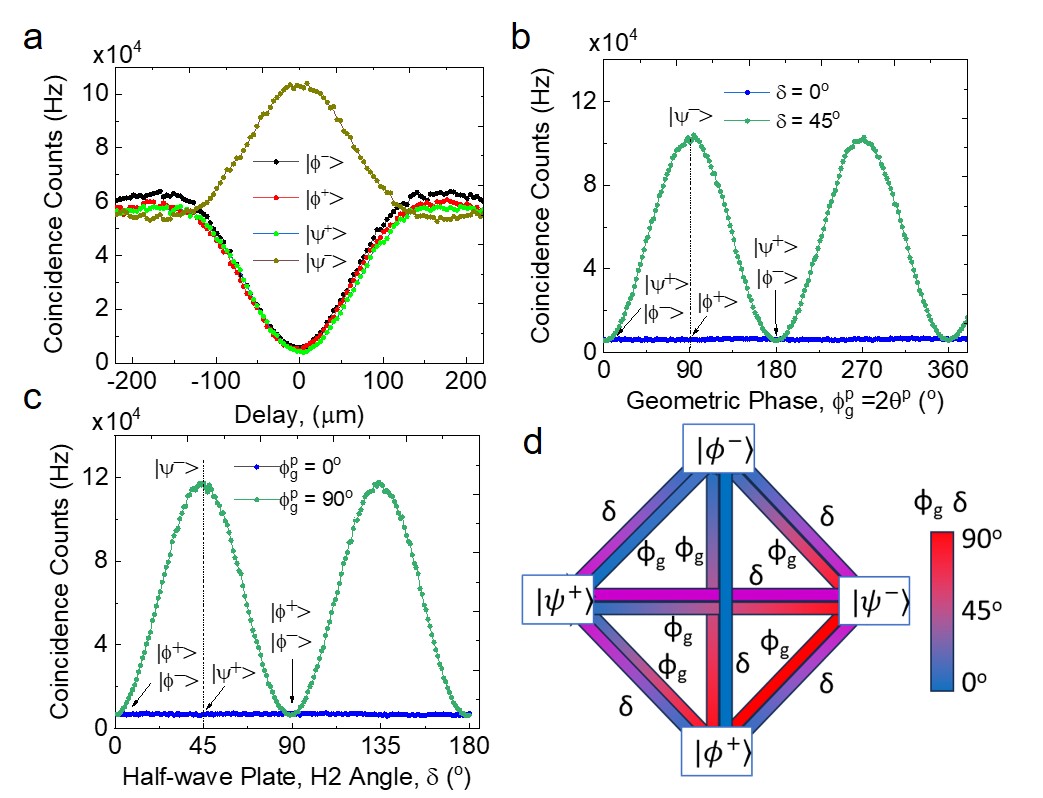}
    \caption{Variation of coincidence counts in HOM interferometry as function of (a) optical delay for four Bell states, (b) relative phase of Bell state due to the geometric phase, and (c) half-wave plate angle for Bell state transformation. (d) Pictorial diagram of Bell state transformation for variation of relative phase and $\delta$.}
   \label{Figure 3}
\end{figure}

After confirming the on-demand flexible generation of all four Bell states, we investigated their HOM interference. The results are shown in Fig. \ref{Figure 3}. Keeping the pump power constant, we prepared different Bell states appropriately setting $\phi_g^p$ and $\delta$ (see Table \ref{tab:bell_results} and Eq. \ref{eq4}), and measured the coincidence counts between the beam splitter (BS) output ports, $c$ and $d$ (see Fig. \ref{Figure 1}, as a function of relative optical delay. As shown in Fig. \ref{Figure 2}(a), for a large delay, the coincidence counts are high for all Bell states. As the delay approached zero, the coincidence counts decreased for the Bell states, $|\Phi^-\rangle$, $|\Phi^+\rangle$, and $|\Psi^+\rangle$, forming the characteristic HOM dip associated with photon bunching. In contrast, for the Bell state $|\Psi^-\rangle$, the coincidence counts peaked at zero delay and decreased symmetrically on either side, producing an anti-dip. These observations are in excellent agreement with Eq. \ref{eq6} and Eq. \ref{eq7}, confirming the symmetric and antisymmetric nature of the generated Bell states. The HOM visibility of the symmetric states is measured to be 91.1$\%$, with a dip width of $\approx124$ $\mu$m. The antisymmetric state exhibited an anti-dip visibility of 95.9$\%$ with a comparable width.

To further examine the role of geometric phase, we fixed the relative delay at zero and varied geometric phase, $\phi_g^p$ for $\delta = 0^\circ$ (blue dots) and $45^\circ$ (green dots), corresponding to the $\ket{\Phi}$ and $\ket{\Psi}$ families, respectively. As shown in Fig. \ref{Figure 3}(b), the coincidence counts (blue dots) remained invariant with $\phi_g^p$ for the $|\Phi^-\rangle$ state, which periodically transformed to $|\Phi^+\rangle$ as $\phi_g^p$ increased from $0^\circ$ to $90^\circ$ and multiples thereof. This behavior is expected since the geometric phase associated with the same-polarization components of the Bell state acts as a global phase, to which HOM interference is insensitive. The relative phase modifies only the internal superposition within the Bell state and does not affect the interference arising from individual terms; therefore, all intermediate phase-shifted Bell states of the $\ket{\Phi}$ family remain symmetric (bosonic-like) Bell states. In contrast, for the $|\Psi^+\rangle$ state (bosonic-like), the coincidence counts exhibit a clear dependence on $\phi_g^p$. As $\phi_g^p$ increases from $0^\circ$ to $90^\circ$, the state evolves continuously into $|\Psi^-\rangle$, passing through intermediate phase-shifted Bell states of the $\ket{\Psi}$ family. This dependence follows a $\sin^2(\phi_g^p)$ variation (see Eq. \ref{eq4}) in the coincidence counts, reaching a fully antisymmetric (fermionic-like) Bell state at $\phi_g^p = 90^\circ$, consistent with Eq. \ref{eq7}. This behavior arises from interference between indistinguishable two-photon probability amplitudes associated with exchange pathways of pair photons in orthogonal polarization states at the beamsplitter, leading to bosonic, anyonic, or fermionic statistics \cite{Sansoni:12} depending on the relative phase of the $\ket{\Psi}$-family Bell states.

We also studied the effect of interchange between the $\ket{\Phi}$ and $\ket{\Psi}$ Bell state families on the HOM interference. Setting the relative optical delay to zero, we used two fixed values of geometric phase, $\phi_g^p$, $0^\circ$ and $90^\circ$, to ensure transitions between $\ket{\Phi^-}$ and $\ket{\Psi^+}$, and between $\ket{\Phi^+}$ and $\ket{\Psi^-}$, depending on the value of $\delta$. As shown in Fig. \ref{Figure 3}(c), for $\phi_g^p = 0^\circ$ (blue dots), the coincidence counts remain unchanged, and for $\phi_g^p = 90^\circ$ (green dots), the coincidence counts vary sinusoidally with $\delta$. This behavior is consistent with Eq. \ref{eq4}. For $\phi_g^p = 0^\circ$, the symmetric state $\ket{\Phi^-}$ transforms into the symmetric state $\ket{\Psi^+}$ at $\delta = 45^\circ$. Intermediate values of $\delta$ produce superpositions of symmetric states, resulting in HOM bunching at zero delay. In contrast, for $\phi_g^p = 90^\circ$, the symmetric state $\ket{\Phi^+}$ transforms into the antisymmetric state $\ket{\Psi^-}$ at $\delta = 45^\circ$. For intermediate $\delta$, the state becomes a superposition of symmetric and antisymmetric components (see Eq. \ref{eq4}), leading to reduced bunching and the emergence of antibunching, with maximum antibunching at $\delta = 45^\circ$ due to the fully antisymmetric state $\ket{\Psi^-}$. The periodic variation of coincidence counts with $\delta$ follows a $\sin^2(2\delta)$ dependence, as also evident from Eq. \ref{eq4}. The transition between the Bell states with variation of the geometric phase $\phi_g^p$, and the angle $\delta$ of the half-wave plate (H2) is pictorially summarized in Fig. \ref{Figure 3}(d). The color bar represents the values of $\phi_g^p$ and $\delta$. The vertices correspond to the four Bell states, while the intermediate points represent phase-shifted Bell states.

It is interesting to note from the green dots of Fig. \ref{Figure 3}(b) and Fig. \ref{Figure 3}(c) that varying the geometric phase, $\phi_g^p$, introduced between the orthogonal polarization components of the pump from $0^\circ$ to $90^\circ$ for a fixed value of $\delta = 45^\circ$, corresponding to the transition pathway from $\ket{\Psi^+}$ to $\ket{\Psi^-}$, or equivalently varying $\delta$ from $0^\circ$ to $45^\circ$ for a fixed value of $\phi_g^p = 90^\circ$, corresponding to the transition pathway from $\ket{\Phi^+}$ to $\ket{\Psi^-}$, results in a large change in the coincidence counts of the order of $\sim 10\times10^4$ counts/s. This observation highlights the potential of the scheme for high-resolution birefringence measurements, either by placing the sample in the pump beam or within one arm of the HOM interferometer, consistent with the analysis of Eq. \ref{eq7}. However, introducing the sample in the classical pump beam is particularly advantageous, since placing the sample in the path of the entangled photons can introduce additional losses, reduce the coincidence count rate, degrade the measurement resolution, and increase the complexity of interferometric alignment.

\begin{figure}[ht]
    \centering
    \includegraphics[width=1\linewidth]{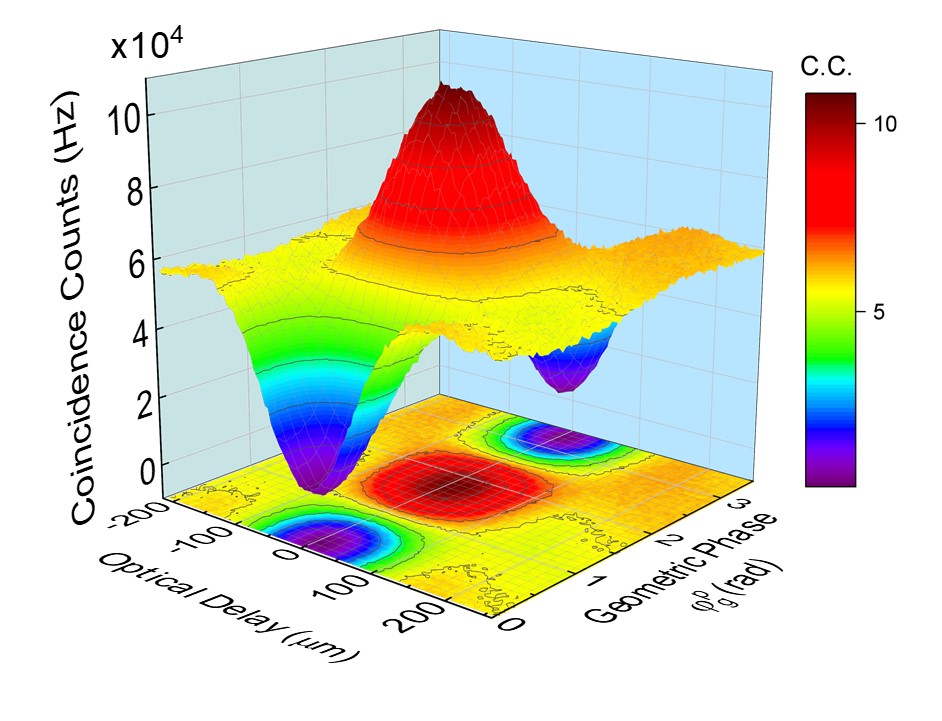}
    \caption{Variation of coincidence counts as function of optical delay in HOM interferometer for different relative phase of the Bell states of $\ket{\Psi}$ family.}
   \label{Figure 4}
\end{figure}

To gain further insight into the sensing capability of the scheme, we used the $\ket{\Psi^+}$ state with $\delta = 45^\circ$ and continuously varied the geometric phase in the pump beam while measuring the HOM interferometry by changing the relative optical delay between the pair photons. The results are shown in Fig. \ref{Figure 4}. As evident from Fig. \ref{Figure 4}, the introduction of a relative phase (geometric phase) in the classical pump beam continuously reduces the HOM interference visibility, reaching zero at $\phi_g^p = \pi/4$ rad. A further increase in the geometric phase reverses the interference behavior, resulting in an increase in the coincidence counts and producing a maximum at $\phi_g^p = \pi/2$ rad, following the periodic dependence of $\sin^2(\phi_g^p)$ predicted by Eq.\ref{eq6},\ref{eq7}. The transition from minimum to maximum coincidence counts is observed for all relative optical delays; however, the modulation range decreases with increasing temporal delay between the photons and becomes maximum at zero optical delay, where the photons are fully indistinguishable. The measured full width at half maximum (FWHM) temporal width of both the HOM dip and anti-dip is determined by the coherence time of the photon pairs and is measured to be \CK{$\approx124$ $\mu$m}, corresponding to the spectral filtering of the photons using a 3 nm bandpass filter. In contrast, the FWHM of the coincidence modulation as a function of the relative phase remains constant at $\pi/2$, independent of the spectral bandwidth of the photon pairs, consistent with Eq. \ref{eq7}. The combined dependence of coincidence counts on both the optical delay and geometric phase is further illustrated in the projection image showing the variation of coincidence counts with relative optical delay and geometric phase.

Having fully characterized the transition from the symmetric (bosonic) to the antisymmetric (fermionic) entangled state, we investigate a proof-of-principle application of this phenomenon for sensing. It is evident from Eq. \ref{eq7} that the coincidence counts vary linearly around $\phi = 45^\circ$, corresponding to the intermediate anyonic-like regime between bosonic and fermionic symmetry. By using the geometric phase as a tunable control parameter to bias the system at this operating point and setting $\delta = 45^\circ$, the HOM interferometer attains its maximum phase sensitivity. Under these conditions, small birefringence-induced phase shifts introduced by a sample placed in the pump beam can be directly mapped onto measurable changes in the coincidence counts.

\begin{figure}[ht]
    \centering
    \includegraphics[width=1\linewidth]{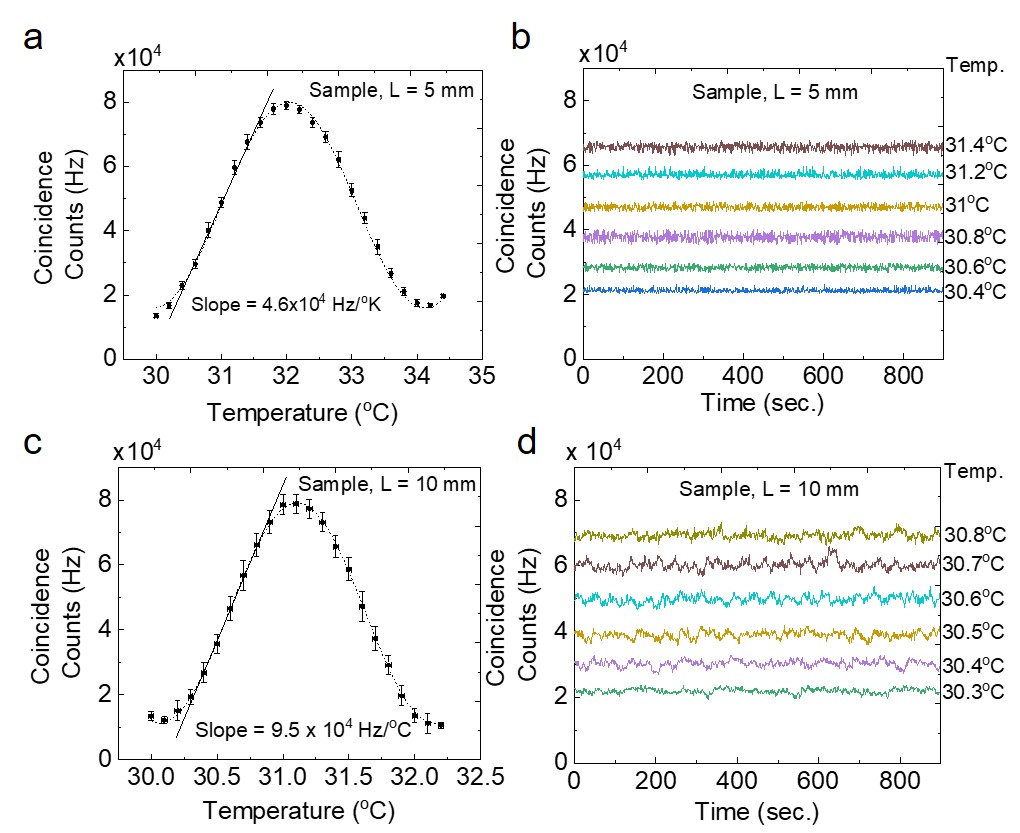}
    \caption{Variation of the coincidence counts as a function of temperature-induced birefringence arising from thermo-optic dispersion for crystals with lengths (a,b) L = 5 mm and (c,d) L = 10 mm. The solid lines represent fits to the experimental data, from which the slopes were determined. Dotted lines are the sinusoidal fit to the experimental data. }
   \label{Figure 5}
\end{figure}

Keeping the experimental setup unchanged, we placed a sample in the pump-beam path after the geometric-phase setup (see Fig. \ref{Figure 1}). Since the relative phase, $\phi$, between the orthogonal polarization components of the pump beam can be introduced through the birefringence of a material, we used a PPKTP crystal with the same aperture (1 $\times$ 2)mm$^2$ but two different lengths, L = 5 mm, and L = 10 mm. Owing to thermo-optic dispersion, the birefringence of the crystal can be tuned by changing its temperature. Therefore, we placed the crystal inside a temperature-controlled oven whose temperature could be varied with an accuracy of ($\pm 0.02^\circ$C). Keeping all other experimental conditions unchanged and adjusting the geometric phase to compensate for the initial birefringence of the crystal at T = $30^\circ$C by ensuring minimum coincidence counts (corresponding to the symmetric state, $\ket{\Psi^+}$), we measured the coincidence counts of the HOM interferometer as a function of crystal temperature. The results are shown in Fig. \ref{Figure 5}. As evident from Fig. \ref{Figure 5}(a), increasing the crystal temperature in steps of ($\Delta T = 0.2^\circ$C results in an increase in the coincidence counts, reaching a maximum value (corresponding to the antisymmetric state, $\ket{\Psi^-}$) at T = $32^\circ$C. Further increases in temperature lead to a periodic variation of the coincidence counts, similar to the behavior observed in Fig. \ref{Figure 3}(b). By fitting a straight line to the linear region of the experimental data shown in Fig. \ref{Figure 5}(a), we obtain a rate of change of coincidence counts as high as 4.6$\times10^4$ counts/s/$^\circ$C for the L = 5 mm long PPKTP crystal. To confirm the reliability of the observed change in coincidence counts with small temperature variations, we measured the stability of the coincidence counts in the linear region of Fig. \ref{Figure 5}(a) for 15 minutes at each temperature step of $\Delta T = 0.2^\circ$C. The results, shown in Fig. \ref{Figure 5}(b), exhibit clearly distinguishable levels separated by approximately 10$^4$ counts/s.
We repeated the same measurements using a crystal of length, L = 10 mm, with the results shown in Fig. \ref{Figure 5}(c) and Fig. \ref{Figure 5}(d). As evident, the coincidence counts vary sinusoidally with crystal temperature, with a transition from minimum to maximum coincidence counts occurring for a temperature change of approximately 1$^\circ$C. The slope of the linear region reaches 9.5$\times$10$^4$counts/s/$^\circ$C, which is nearly twice that obtained for the L = 5 mm crystal, consistent with the doubling of the crystal length. The enhanced sensitivity also enables the resolution of birefringence changes corresponding to temperature variations as small as $\Delta T = 0.1^\circ$C, as demonstrated by the distinct count levels separated by approximately 10$^4$ counts/s in Fig. \ref{Figure 5}(d). Since the sensitivity increases with crystal length, the finite temperature fluctuation of the oven ($\pm$0.02$^\circ$C) produces larger fluctuations in the measured coincidence counts, which are visible in Fig. \ref{Figure 5}(d) compared with Fig. \ref{Figure 5}(b). Owing to the unavailability of a temperature controller with higher precision, we were unable to characterize the temperature-dependent birefringence for even smaller temperature changes despite the high intrinsic resolution of the sensing scheme. By comparing the results of Fig. \ref{Figure 5}(a) and Fig. \ref{Figure 5}(c) with the periodic behavior observed in Fig. \ref{Figure 3}(b), we estimate that the birefringent phase of the PPKTP crystal at 405 nm changes by approximately 90$^\circ$ for a temperature variation of 2$^\circ$C and 1$^\circ$C for crystal lengths of L = 5 mm and L = 10 mm, respectively, around T = 30$^\circ$C.

The phase difference between the orthogonal polarization components is given by, $\phi$ = (2$\pi/\lambda)\times \Delta n(T) L(T)$, where $\Delta n(T)$ = $n_e(T)$ - $n_o(T)$ is the birefringence arising from thermo-optic dispersion and L(T) accounts for the change in crystal length due to thermal expansion characterized by the coefficient ($\alpha$). Using the thermo-optic dispersion equations reported in Ref. \cite{Kato:2002} and the thermal expansion coefficient from Ref. \cite{Emanueli:2003}, we calculate the temperature-dependent birefringent phase sensitivity of PPKTP for $x$-$z$ and $y$-$z$ polarization states in the wavelength range 850–530 nm, where the thermo-optic dispersion equations are valid. The calculated phase sensitivities vary from 41.8–108$^\circ/^\circ$C and 31.9–81$^\circ/^\circ$C for the $x$-$z$ and $y$-$z$ polarization pairs, respectively. Since the thermo-optic dispersion equations of Ref. \cite{Kato:2002} are not valid at 405 nm, extrapolation of the formulas to this wavelength yields unusually large values and should be interpreted with caution. Nevertheless, the experimental results clearly demonstrate that the proposed scheme is capable of resolving birefringence changes on the order of $10^{-6}–10^{-7}/^\circ$C for a 10 mm birefringent sample (here PPKTP). Furthermore, the use of geometric phase as a controllable pointer allows the operating point of the sensor to be reset to the region of maximum sensitivity. Combined with compensation of the birefringence-induced phase shift with geometric phase, this approach can provide high-resolution measurements over a broad dynamic range \cite{Singh:24}.

\begin{figure}[ht]
    \centering
    \includegraphics[width=0.9\linewidth]{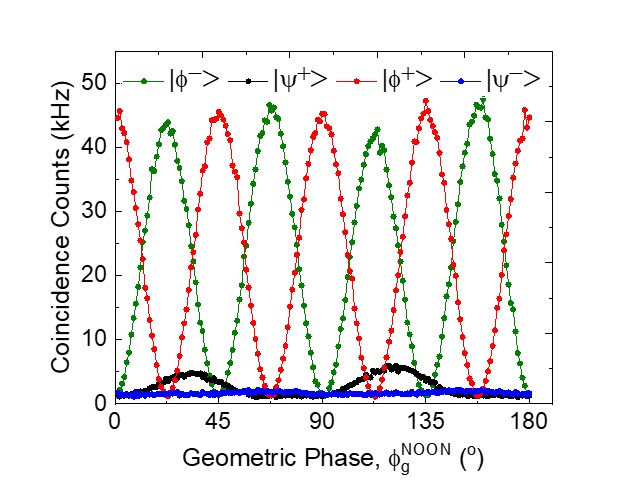}
    \caption{Variation of coincidence counts with a function of the relative phase of the NOON states generated from maximally entangled Bell states.}
   \label{Figure 6}
\end{figure}

We further verified the generation of the NOON state as predicted by Eq. \ref{eq5} for different combinations of $\phi_g^p$ and $\delta$ presented in Table \ref{tab:bell_results}. Using the flip mirror (FM), we have reflected the photons of the output port $c$ of the beam splitter and measured the coincidence counts between the two output ports of the PBS while changing the geometric phase $\phi_g^{NOON}$ by the waveplate, H3. The results are shown in Fig. \ref{Figure 6}. As expected from Eq. \ref{eq5}, the NOON states generated for Bell states of $\ket{\phi}$ family (red and green dots and lines) show coincidence counts as a function of the $\phi_g^{NOON}$ introduced to orthogonal polarization states with twice the period due to the localization of two photons in the same mode. On the other hand, the NOON states generated for Bell states of $\ket{\Psi}$ family (black and blue dots and lines) show coincidence counts independent of the $\phi_g^{NOON}$ introduced to orthogonal polarization states. The small fluctuation in the coincidence counts for the NOON state formed by the $\ket{\Psi^+}$ can be attributed to the residual phase of the waveplate used in the experiment. Therefore, one can use the $\ket{\phi}$ family Bell states for the NOON state based quantum enhanced multiple phase estimation with high resolution \cite{Dowling:2008, Humphreys:2013}.

\section{Conclusions}
In conclusion, we have demonstrated a quantum sensing scheme based on Hong–Ou–Mandel interference that exploits controlled transitions between the symmetric and antisymmetric Bell states. Unlike conventional interferometric approaches that rely on first-order coherence and phase-stable optical paths, the present method extracts phase information from two-photon coincidence measurements arising from changes in the symmetry of an entangled state. By introducing a controlled geometric phase in the pump beam, we achieve a tunable modulation of the coincidence signal through the conversion between the $\ket{\Psi^{+}}$ and $\ket{\Psi^{-}}$ Bell states, enabling the measurement of minute phase variations. As a proof of principle, we employed a birefringent PPKTP crystal as the sensing element and measured its thermo-dispersive birefringence with a resolution on the order of $10^{-6}$ over a broad temperature range. Furthermore, we demonstrated the use of geometric phase as a controllable cursor to operate the sensor within its high-sensitivity linear regime and to extend its dynamic range by resetting the coincidence response to its initial operating point. The measured response exhibits a high signal-to-noise ratio, intrinsic resilience to background noise, and robustness against common-mode phase fluctuations. Although the normalized phase sensitivity is comparable to that of a classical polarization interferometer measuring the same relative phase, the sensing mechanism relies on a genuinely quantum resource, the controlled manipulation of entangled-state symmetry, without any classical analogue. These results establish Bell-state symmetry transitions as a powerful and practical resource for quantum metrology and open new avenues for robust quantum sensing based on structured light, spin–orbit interactions, and high-dimensional photonic states.

\begin{acknowledgments}
The authors acknowledge the support of the Department of Space, Govt. of India.
\end{acknowledgments}

\section*{Data availability} Data underlying the results presented in this paper are not publicly available at this time but may be obtained from the authors upon reasonable request.
\section*{Disclosures}
The authors declare no conflicts of interest.

\nocite{*}
\bibliographystyle{unsrt}  
\bibliography{References}  

\end{document}